\documentclass[%
 reprint,
 superscriptaddress,
 amsmath,amssymb,
 aps,
 pra,
floatfix
]{revtex4-1}

\usepackage[paperwidth=210mm,paperheight=297mm,centering,hmargin=2.3cm,tmargin=2.8cm,bmargin=3.4cm]{geometry}
\usepackage{booktabs}
\usepackage{subfigure}
\usepackage{graphicx,epsf}
\usepackage{amsfonts}
\usepackage{amssymb}
\usepackage{amsmath}
\usepackage{overpic}
\usepackage{braket}
\usepackage{xcolor}
\usepackage{cancel}
\usepackage{fancyhdr}
\usepackage[normalem]{ulem}

\newcommand{\be}{\begin{equation}}
\newcommand{\ee}{\end{equation}}
\newcommand{\bea}{\begin{eqnarray}}
\newcommand{\eea}{\end{eqnarray}}

\def\lb{\label}

\newcount\bozza \bozza=1
\ifnum\bozza=1
\newdimen\shift \shift=-2truecm
\def\lb#1{%
{\label{#1}\rlap{\kern\shift{$\scriptstyle#1$}}}}
\else\def\lb#1{\label{#1}} \fi

\allowdisplaybreaks

\begin{document}

\title{Uniformly frustrated XY model: strengthening of the vortex lattice by intrinsic disorder}

\author{I. Maccari}
\affiliation{Department of Physics, Stockholm University, 106 91 Stockholm, Sweden}
\author{L. Benfatto}
\affiliation{ISC-CNR, Department of Physics, Sapienza University of Rome, P. le A. Moro 5, 00185 Rome, Italy}
\author{C. Castellani}
\affiliation{ISC-CNR, Department of Physics, Sapienza University of Rome, P. le A. Moro 5, 00185 Rome, Italy}

\begin{abstract}
In superconducting films, the role of intrinsic disorder is typically to compete with superconductivity by fragmenting the global phase coherence and lowering the superfluid density. Nonetheless, when a transverse magnetic field is applied to the system and an Abrikosov vortex lattice forms, the presence of disorder can actually strengthen the superconducting state against thermal fluctuations. By means of Monte Carlo simulations on the uniformly frustrated XY model in two dimensions, we show that while for weak pinning the superconducting critical temperature $T_c$ increases with the applied field $H$, for strong enough pinning the experimental decreasing dependence between $T_c$ and $H$ is recovered with a resulting more robust vortex lattice.
\end{abstract}
\maketitle

\section{Introduction}

The melting transition of two-dimensional (2D) Abrikosov vortex lattice (VL) has always attracted a significant experimental and theoretical interest. 
The interplay between magnetic field, random pinning and phase fluctuations makes the phase diagram of the system rich of different phases of matter ranging from the 2D Bragg glass to the 2D vortex glass~\cite{fisherThermalFluctuationsQuenched1991, blatterVorticesHightemperatureSuperconductors1994, kleinBraggGlassPhase2001}, from the isotropic vortex liquid to the more recently observed hexatic liquid phase~\cite{guillamonDirectObservationMelting2009, royMeltingVortexLattice2019}. 
In the absence of disorder, at low temperatures and low vortex densities, the vortex lattice exhibits a quasi-long range order. As the temperature or the vortex density increases, topological defects such as dislocations and disclinations form, eventually leading to the melting of the vortex lattice according to the Berezinskii-Kosterlitz-Thouless theory~\cite{berezinskyDestructionLongrangeOrder1972, kosterlitzOrderingMetastabilityPhase1973, kosterlitzCriticalPropertiesTwodimensional1974} afterwards refined by Nelson, Halperin and Young~\cite{halperinTheoryTwoDimensionalMelting1978, nelsonDislocationmediatedMeltingTwo1979, youngMeltingVectorCoulomb1979}.
In this frame, the role of disorder is on the one hand to favour the formation of such a lattice defects turning the ground state from a quasi ordered Bragg glass~\cite{giamarchiElasticTheoryFlux1995, ledoussalDislocationsBraggGlasses2000} to a disordered vortex glass. On the other hand, by acting as a pinning potential for vortices it also prevents the VL from sliding throughout the system and destroying the global superconducting (SC) phase coherence.
Thus, differently from the Meissner state, where disorder competes with superconductivity, in the mixed state its role is less straightforward. Here, indeed, a true superconducting state exhibiting zero resistivity in the limit of zero current can only settle if certain degree of disorder is present within the system. Moreover, as recently shown~\cite{gangulyMagneticFieldInduced2017}, vortex thermal fluctuations responsible for the fast depletion of the superfluid stiffness can be even reduced by the presence of a strong vortex pinning.

From a theoretical perspective, one possible way to investigate such interplay is by studying the classical XY model in the presence of both disorder and transverse magnetic field. 
As an effective model for superconducting phase fluctuations, the 2D XY model has been over the years applied to the study of disordered and inhomogeneous SC films~\cite{barabashConductivityDueClassical2000, wysinExtinctionBerezinskiiKosterlitzThoulessPhase2005,erezEffectAmplitudeFluctuations2013, costaKosterlitzThoulessTransitionDiluted2014, maccariNumericalStudyPlanar2016, kumarOrderingKineticsRandombond2017, maccariBroadeningBerezinskiiKosterlitzThoulessTransition2017,maccariDisorderedXYModel2019} as well as to the 2D vortex lattice melting for different values of the applied magnetic field~\cite{teitelPhaseTranstionsFrustrated1983, franzVortexLatticeMelting1994, franzVortexlatticeMeltingTwodimensional1995, hattelFluxlatticeMeltingDepinning1995, tanakaNumericalStudyFlux2001, hasenbuschMulticriticalBehaviourFully2005, albaUniformlyFrustratedTwodimensionalXYmodel2008, teitelTwoDimensionalFullyFrustrated2013}.
The combined effect of quenched disorder and transverse magnetic field has been so far addressed by including in the XY model spatially random quenched magnetic fluxes, but with a resulting zero net magnetic field~\cite{blatterVorticesHightemperatureSuperconductors1994, barabashConductivityDueClassical2000,albaQuasilongrangeOrder2D2009, albaMagneticGlassyTransitions2010}. 

In this manuscript, we will instead consider the effect of quenched disorder on the melting of the 2D vortex lattice which forms when a nonzero net transverse magnetic field is included within the XY model. 
%
Starting from the consideration that the presence of a lattice introduces a pinning effect at low temperature even in the absence of disorder, our Monte Carlo simulations show that apart from disordering the VL ground state, the presence of inhomogeneities makes the vortex lattice stronger against thermal fluctuations with respect the homogenous case according with the experimental observations~\cite{gangulyMagneticFieldInduced2017}.  
The presence of a very week pinning can indeed lead to a very counter-intuitive observation, namely  the increase of the SC critical temperature by the increase of the applied magnetic field. However, when the disorder is sufficiently strong the experimentally observed dependence between $T_c$ and $H$ is recovered. 


 
\section{Materials and Methods}

%
%

We consider a two-dimensional XY model in the presence of random pinning and transverse magnetic field. Its Hamiltonian read:

\begin{equation}
H_{XY}= -\sum_{\mu=\hat{x}, \hat{y}, i} J_{i}^{\mu} \cos(\theta_i - \theta_{i+\mu} + F^{\mu}_i),
\label{Hxy_A}
\end{equation}
\noindent
where $\theta_i$ is the SC phase at site $i$ of a $L_x\times L_y$ lattice, $J_{i}^{\mu}$ are the random couplings between the neighbouring sites $i$ and $i+\mu$ and the phase shift $F^{\mu}_i$ accounts via the minimal substitution for the presence of a finite transverse magnetic field $H\hat{z}=  \vec{\nabla} \times \vec{A}$. Being:

\begin{equation}
\label{Fmui}
F^{\mu}_i= \frac{2 \pi}{\Phi_0} \int_{r_i}^{r_{i+\mu}} A_i^{\mu} \cdot d\mu,
\end{equation}
\noindent
with $\Phi_0= hc/2e$ the quantum flux, and $A_i^{\mu}$ is the vector potential along the bond connecting two neighbours spins $i$ and $i+\mu$. The sum of $F^{\mu}_i$ going counterclockwise around any closed path $C$ of bonds on the lattice is $2\pi$ times the number of magnetic flux quanta $f_C$ penetrating the path:
\begin{equation}
\sum_{C}F^{\mu}_i= \frac{2 \pi}{\Phi_0}\oint_C \vec{A} \cdot \vec{dl} = 2 \pi \frac{\Phi_C}{\Phi_0}= 2\pi f_C
\end{equation}
\noindent
Since in the following we will always consider $H$ to be uniform in space, we will refer to the intensity of the applied field in terms of the flux quanta $f$ penetrating through an unitary plaquette $P$:

\begin{equation}
2\pi f=\frac{2 \pi}{\Phi_0}\oint_P \vec{A} \cdot \vec{dl}=\frac{2 \pi}{\Phi_0}\oint_P \vec{\nabla}\times \vec{H} \cdot \vec{dl}=\frac{2\pi}{\Phi_0} H a^2
\end{equation}
with $a=1$, so that $f=\frac{H}{\Phi_0}$.
In literature , one usually refers to this case as the \emph{uniformly frustrated} XY model with frustration $f$. 
Indeed, the phase shift $F^{\mu}_i$ in the cosine argument of \eqref{Hxy_A} adds \emph{frustration} to the system by rendering the ground state no longer ferromagnetic: at $T=0$, the phases $\theta_i$'s instead to be all equal, will vary from site to site trying to minimise the new gauge-invariant phase $(\theta_i - \theta_{i+\mu} + F^{\mu}_i)$. Consequently, the value of $f$ will correspond to the level of such frustration, determining the inhomogeneous space structure of the ground state itself.
Specifically, the ground state of the uniformly frustrated XY model will consist of a periodic configuration of vortices in the phase angle $\theta_i$, whose number is directly proportional to $f$.
The number of vortices of the ground state for a given value of $f$, can be easily derived by rewriting the charge neutrality condition in terms of the new gauge invariant phase: 
\begin{equation}
\sum_j \oint_{P_j}( \vec{\nabla} \theta_i -\vec{A}_i)= 2 \pi \sum_i (n_i-f)=2 \pi \sum_i q_i=0 
\label{charge_cons_H}
\end{equation}
so that: 

\begin{equation}
N_v=fL^2
\label{vortices_f}
\end{equation} 
Since in the present study we will consider periodic boundary conditions, a good gauge choice for the vector potential $\vec{A}$ is the Coulomb gauge: $ \vec{\nabla}\cdot\vec{A}=0$. For simplicity, we will consider:

\begin{equation}
\vec{A}= B(0, x),
\end{equation} 
\noindent
so that:

\begin{equation}
F^{\mu}_i=
\begin{cases}
0 &\text{ if $\mu =x$}\\
2 \pi f x_i &\text{ if $\mu=y$}
\end{cases}
\label{gauge_f}
\end{equation}

\noindent
Finally, it is important to highlight that with this choice not all the values of $f$ will be allowed. Indeed, periodic boundary conditions, together with the Coulomb gauge $A_y=2\pi f x$, give rise to the constraint:
\begin{equation}
L_x \cdot f= 1, \, 2, \, 3, \, \dots
\label{f_constraint}
\end{equation}
\noindent
Hence, for a given value of $L_x$ the smallest frustration we can introduce within the system is: $f= 1/L_x$.\\
In the present work, we have studied the model Eq.\eqref{Hxy_A} on a square lattice with periodic boundary conditions for a linear size of $L_x=L_y=L=64$ and different values of $f$.
In our Monte Carlo (MC) simulations, we have used a local Metropolis algorithm, needed to probe the correct canonical distribution of the system, combined with a micro-canonical Over-relaxation algorithm. Specifically, each Monte Carlo step consists of $5$ Metropolis spin flips of the whole lattice, followed by $10$ Over-relaxation sweeps of all the spins. To help the correct thermalization of the system at lower temperatures, we have used a Simulated-Annealing procedure. For each run we have made 50000-75000 MC steps, measuring the main observables with a frequency of 5 steps, after having discarded the firsts 25000.  Finally, the averages have been computed over 5 independent runs for the clean case and over 10 different realization of quenched disorder for the disordered case.

In the present work, together with the ground state of the vortex lattice, we have studied the SC transition as function of the frustration $f$ and the level of disorder. To address this issue and measure the SC phase coherence of the system, we have computed the superfluid stiffness $J_s^{\mu}$ which accounts for the linear response of the system to an infinitesimal twist $\Delta_{\mu}$ of the gauge invariant phase along a given direction $\mu$.  As such, it is 
defined as the second derivative of the free energy with respect to $\Delta_{\mu}$ at $\Delta_{\mu}=0$: $$J_s^{\mu}\equiv - \frac{1}{L^2}\frac{\partial^2 \ln{Z(\Delta_{\mu})}}{\partial \Delta_{\mu}^2}\large|_{\Delta_{\mu}=0}$$
being it finite in the SC phase and zero in the normal phase.
Its expression for the model Eq.\eqref{Hxy_A} reads: 

\begin{equation}
\begin{split}
J_s^{\mu}= \frac{1}{L^2}&\big\langle \sum_i J_i^{\mu} \cos( \theta_i - \theta_{i + \mu} +F^{\mu}_i) \big\rangle +\\
 -&\frac{\beta}{L^2} \Bigg[\big\langle \left( \sum_i J_i^{\mu} \sin(  \theta_i - \theta_{i + \mu} +F^{\mu}_i) \right)^2  \big\rangle +\\ -& \big\langle \sum_i J_i^{\mu} \sin(  \theta_i - \theta_{i + \mu} +F^{\mu}_i)\big\rangle^2 \Bigg],
\end{split}
\label{Js_magn}
\end{equation}
where $\beta$ is the inverse temperature and $\langle \dots \rangle$ stays both for the  Monte Carlo thermal average and for the average over the independent runs.

\section{Results}

In what follows, we will present the numerical results obtained via Monte Carlo simulations both for the clean and for the disordered case, considering different values of $f \in [0, \frac{1}{2}]$.
\subsection{Clean case}

Let us start with the clean case where $J_i^{\mu}=1$ $\forall i,\mu$ so that:
\begin{equation}
H_{XY}= -J \sum_{\mu=\hat{x}, \hat{y}, i} \cos(\theta_i - \theta_{i+\mu} + F^{\mu}_i).
\label{Hxy_clean}
\end{equation}
For $f=0$, the model \eqref{Hxy_clean} is the classical XY model which undergoes to the well known Berezinskii-Kosterlitz-Thouless~\cite{berezinskyDestructionLongrangeOrder1972, kosterlitzOrderingMetastabilityPhase1973, kosterlitzCriticalPropertiesTwodimensional1974} (BKT) transition at $T=T_{BKT}$. As already mentioned, for finite values of $f$ the phase transition is no longer driven by the unbinding of vortex-antivortex pairs, but rather by the melting of the vortex lattice which naturally forms when a transverse magnetic field is applied to the system. 
In the clean case, despite the absence of disorder, the vortex lattice is pinned at low temperatures by the presence of the underlying square grid that defines the array of Josephson junctions. 
By acting as a periodic pinning potential for vortices, such a square grid can become particularly relevant for large values of $f$ eventually determining the symmetry of the vortex lattice itself.  
The most emblematic example is the limiting case $f=1/2$, where in the ground state the vortex lattice assumes a checkerboard ordered pattern. 
The model Eq.\eqref{Hxy_clean} with $f=1/2$ is also known as the \emph{fully frustrated} XY model (FFXY) and as such it has been extensively studied (see ~\cite{teitelTwoDimensionalFullyFrustrated2013} and reference therein) with particular focus on its critical behaviour and the nature of the phase transitions it undergoes. 

\begin{figure*}
\centering
\includegraphics[width= 0.9 \linewidth]{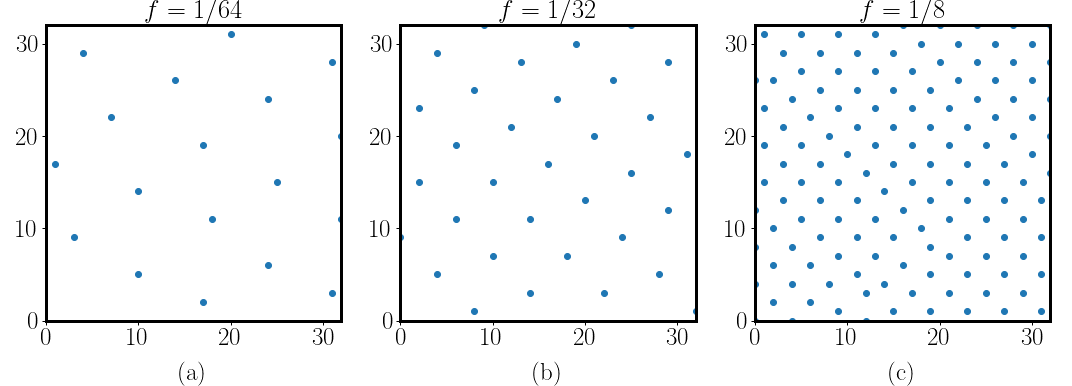}
\caption{Vortex lattice ground state for different values of the frustration $f=1/8;\, 1/32;\, 1/64$. Each vortex core is plotted as a blue point. The linear size of the simulated system is $L=64$, while for the sake of clarity in the figure only a portion of the whole VL is shown.  \label{fig1}}
\end{figure*}

In the present work, we will instead focus on smaller value of $f$, where the periodic pinning potential does not induce such peculiar vortex lattice configuration. The resulting vortex lattice found numerically in the ground state are shown in Fig.\ref{fig1}. For all the values of $f$ considered, at low temperatures the vortices form an ordered and pinned lattice whose symmetry partially depends also on the commensurability with the underlying numerical grid.

\begin{figure}
\centering
\includegraphics[width= \linewidth]{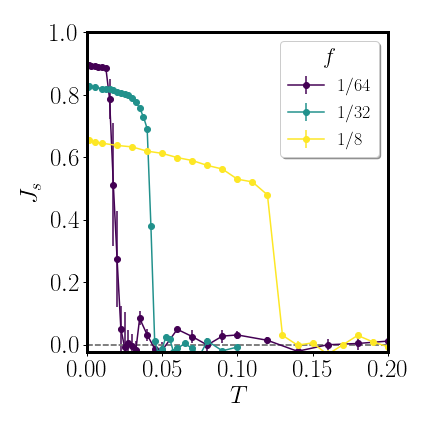}
\caption{Superfluid stiffness $J_s$ as a function of the temperature $T$ for different values of frustration $f$ in the absence of disorder. The linear size of the simulated system is $L=64$.\label{fig2}}
\end{figure}

At finite but not maximal $f$ different kind of phase transitions can occur~\cite{franzVortexlatticeMeltingTwodimensional1995, albaUniformlyFrustratedTwodimensionalXYmodel2008}, and their nature is still unclear in most of the cases.
Without pretending to address this issue, in the present work we focus on the dependence of the critical temperature $T_c$, at which the superfluid stiffness vanishes, on the applied magnetic field $f$. The numerical results for different values of $f$ are reported in Fig.\ref{fig2}.  

We can immediately notice that the critical temperature $T_c$  at which the system loses its phase coherence is strongly suppressed for smaller values of $f$, completely at odd with the usual experimental observations~\cite{yazdaniObservationKosterlitzThoulesstypeMelting1993, blatterVorticesHightemperatureSuperconductors1994, feigelmanVorticesHightemperatureSuperconductors1994, chenTwodimensionalVorticesSuperconductors2007, guillamonDirectObservationMelting2009, misraMeasurementsMagneticFieldTunedConductivity2013, benyaminiFragilityDissipationlessState2019, royMeltingVortexLattice2019}.
The general trend seems, indeed, to be a proportionality between $T_c$ and $f$: lower critical temperatures for smaller frustration.\\  The observed trend has been already discussed in~\cite{albaUniformlyFrustratedTwodimensionalXYmodel2008}, where Alba et al. have shown that in the limit of small frustration $f=1/n$ (and $n \gg 1$), the critical temperature $T_c$ decreases with the increase of $n$: $T_c \sim 1/n \to 0$ as $n\to \infty$.

{Apart from specific cases where the commensurability with the underlying square grid particularly strengthens or weakens the vortex-lattice structure, by lowering the applied magnetic field the pinning of the vortex lattice becomes weaker than in the case where the vortex density is higher, with a resulting decrease of the critical temperature with $f$.    
By increasing the pinning potential via the introduction of disorder, however, the scenario is reversed and, in agreement with experimental observations, with increasing applied magnetic field the critical temperature decreases.
We will discuss the numerical results of the disordered case in the following section.

\subsection{Disordered case}
We now consider the case of a transverse magnetic field applied to a disordered SC film.
In the present manuscript, we will use as disordered coupling constants $J_i^{\mu}$ in Eq.\eqref{Hxy_A} the inhomogeneous local stiffness derived from the (quantum) XY pseudo-spin $1/2$ model in random transverse field (RTF)~\cite{maLocalizedSuperconductors1985, ceaOpticalExcitationPhase2014}.
This disorder has been shown to be appropriate to model disordered superconductors with a non-trivial spatial structure~\cite{ioffeDisorderDrivenQuantumPhase2010, lemarieUniversalScalingOrderparameter2013, ceaOpticalExcitationPhase2014}, as well as to account for the experimental observation of a rather broad BKT transition around the critical temperature $T_{BKT}$~\cite{maccariBroadeningBerezinskiiKosterlitzThoulessTransition2017, ilariamaccariBKTUniversalityClass2018}, at which a sharp jump of the superfluid-density would be expected for zero disorder~\cite{nelsonUniversalJumpSuperfluid1977}. Previous studies at zero magnetic field\cite{maccariBroadeningBerezinskiiKosterlitzThoulessTransition2017, ilariamaccariBKTUniversalityClass2018} have shown that such a spatially-correlated disorder with large enough low-stiffness puddles is crucial to induce an anomalous vortex nucleation and, consequently, a smearing of the BKT transition. On the other hand in the present case where vortices are induced by a finite transverse magnetic field we do not expect that the results will depend crucially on the choice of disorder, so our finding should be general also for different disorder realizations. The level of disorder is here labeled by $W/J$ (see~\cite{maccariBroadeningBerezinskiiKosterlitzThoulessTransition2017} for more details) and it is taken to be quenched in temperature. 

Let us start by considering a relatively weak disorder level $W/J=4$. 
Consistently with the experimental observations of~\cite{gangulyMagneticFieldInduced2017}, our  MC numerical results reveal that the presence of disorder leads to a modification of the ground-state vortex lattice,  enlightening further its underlying mechanism. As shown in Fig.\ref{fig3}, the core of the vortices is indeed pinned by the inhomogeneity of the local stiffness, which makes them moving towards those regions with lower ${J}_i^{\mu}$ so to gain in energy by minimising the Hamiltonian\eqref{Hxy_A}.
To highlight the correlation in space between low-couplings regions and the vortex lattice deformation, in Fig. \ref{fig3} we have superimposed the vortex lattice to the couplings map, obtained by computing over each plaquette the average value of the local stiffness $J^{\mu}_i$.

The presence of disorder not only modifies the vortex lattice ground state, but also impact the superfluid response of the system as function of the applied magnetic field. 
In Fig.\ref{Js_WJ4_f}, we report the temperature dependence of the superfluid stiffness $J_s(T)$ obtained for different values of $f$.
As one can see, the presence of disorder restores the experimentally measured dependence between $T_c$ and $f$ by rendering more robust the vortex lattice against thermal fluctuations. In this regard, it is quite impressive to notice that, compared to the homogeneous case (see Fig.\ref{fig3}), the critical temperature $T_c$ corresponding to the lowest value of $f=1/64$ has increased by a factor of ten by the effects of the inhomogeneity.

The strengthening of the vortex lattice due to disorder is even more pronounced when looking at stronger disorder regimes.
In Fig.\ref{Js_WJ10_f}, we report the superfluid stiffness trend in temperature for the same values of $f$ but with a disorder level of $W/J=10$. Looking for instance at the lowest value of the field ($f=1/64)$, the critical temperature is reduced, because of the field, just by half with respect to the zero-field value, while at weak disorder ($W/J=4$) it was five times smaller.

\begin{figure}
\centering
\includegraphics[width= \linewidth]{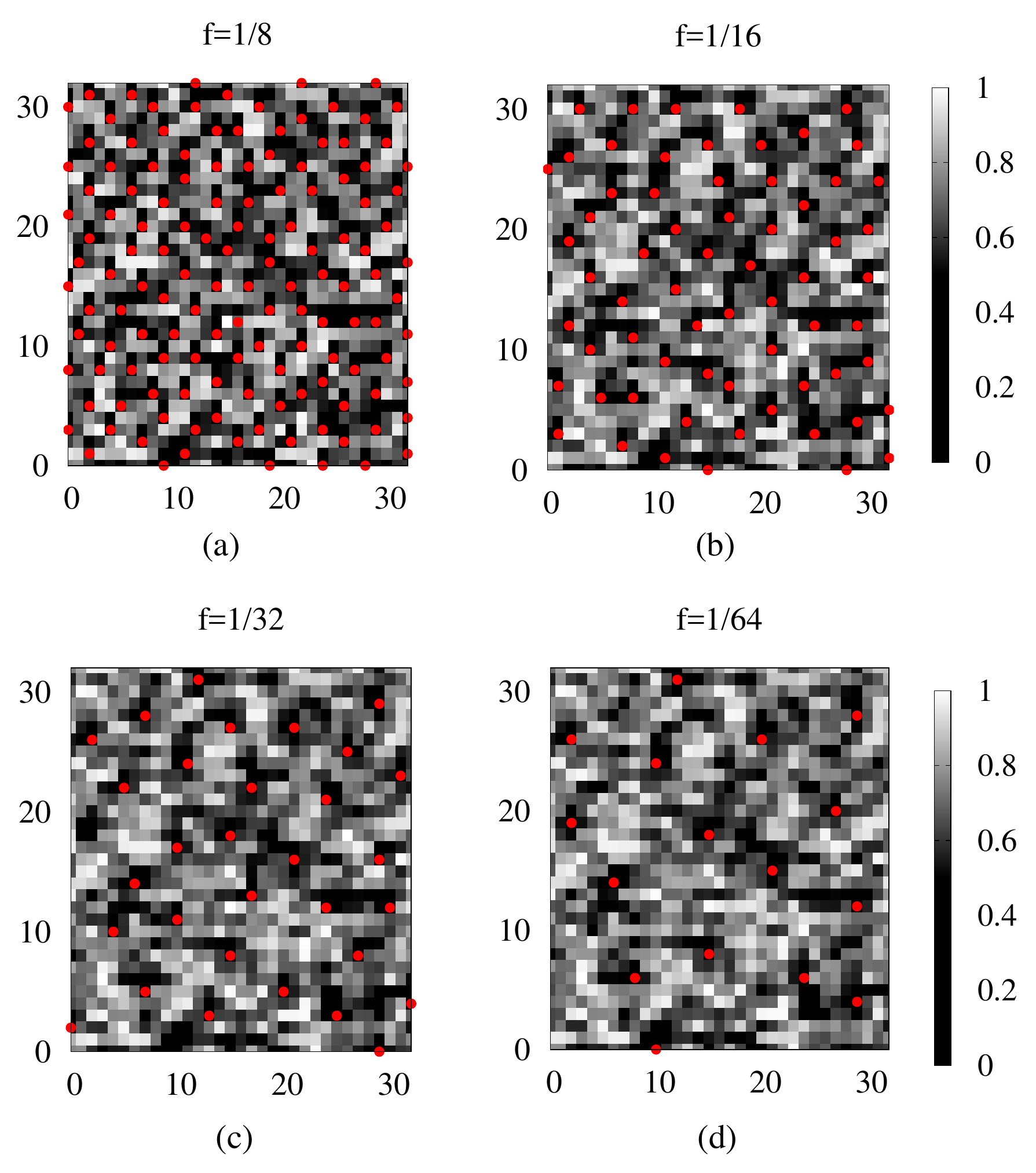}
\caption{Vortex lattice ground state for different values of the frustration $f= 1/8;\, 1/16;\, 1/32;\, 1/64$ superimposed to a map of a given disorder realization with $W/J=4$. The gray scale refers to the the average value of the local couplings constant $J
_i^{\mu}$ around each single plaquette. Each vortex core is plotted as a red point. The linear size of the simulated system is $L=64$, while for the sake of clarity in the figure only a portion of the whole VL is shown.  \label{fig3}}
\end{figure} 

\begin{figure}
\centering
\includegraphics[width= 0.8\linewidth]{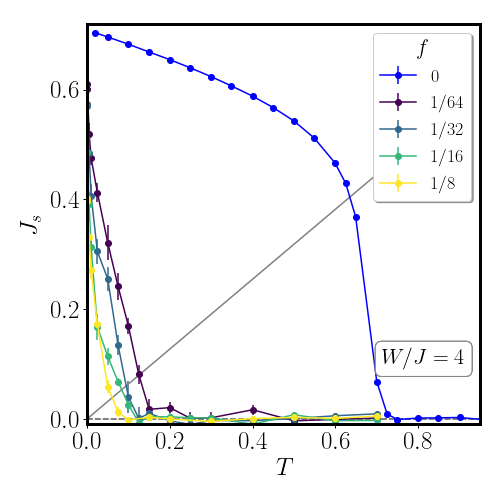}
\caption{Temperature dependence of the superfluid stiffness in the presence of weak disorder $W/J=4$, for different values of the frustration $f$. The continuous gray line is the BKT critical line $2T/\pi$ relative to the case $f=0$. \label{Js_WJ4_f}}
\end{figure}

\begin{figure}[h!]
\centering
\includegraphics[width= 0.8\linewidth]{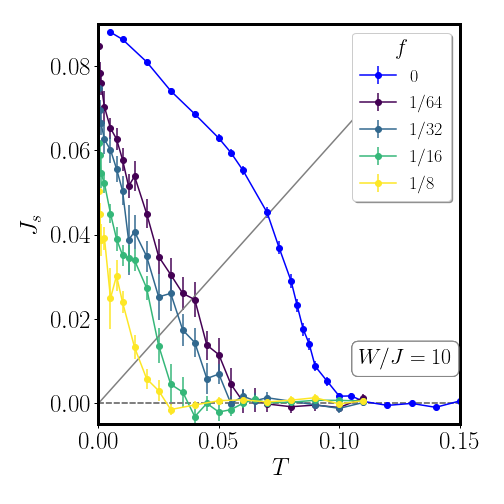}
\caption{Temperature dependence of the superfluid stiffness in the presence of strong disorder $W/J=10$, for different values of the frustration $f$.The continuous gray line is the BKT critical line $2T/\pi$ relative to the case $f=0$.\label{Js_WJ10_f}}
\end{figure}   

In order to highlight such increase of robustness, as effect of the increase of the intrinsic disorder, we have reported in Fig.\ref{fig6} (a) the extrapolated values of the critical temperature as function of the applied field, for the two disorder regimes considered. In Fig.\ref{fig6}(b), this effect is made even more evident by rescaling the curves of $T_c$ by their value in absence of magnetic field.

\begin{figure*}[ht!]
\centering
\includegraphics[width=\linewidth]{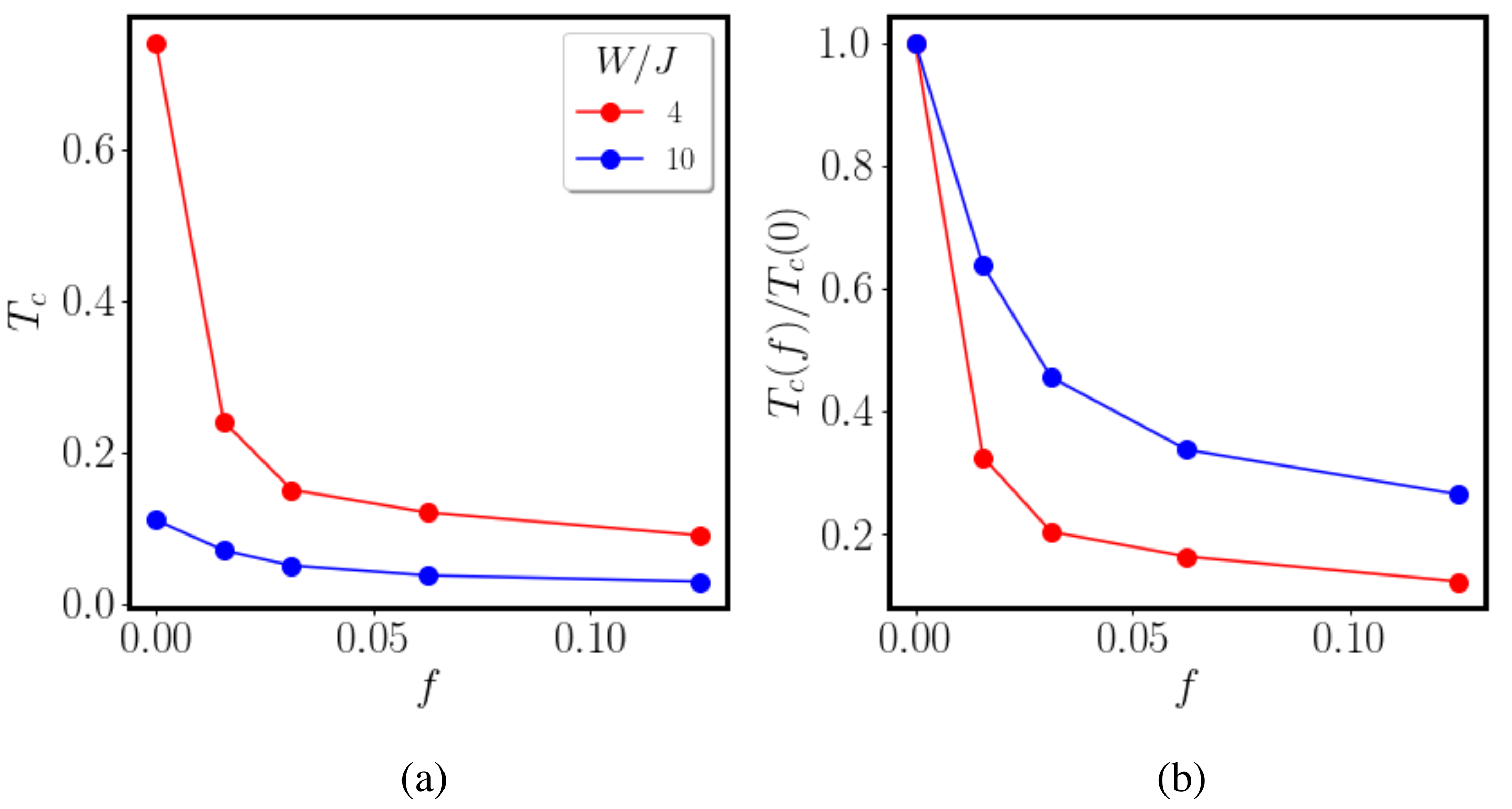}
\caption{Magnetic field dependence of (a) the critical temperature $T_c(f)$ and (b) of  the rescaled critical temperature $T_c(f)/T_c(f=0)$ for the two regime of disorder considered: $W/J=4$ and $W/J=10$. For each values of $f\neq0$ and disorder level, the critical temperature $T_c$ has been estimated from the superfluid stiffness trend in temperature $J_s(T)$ as the temperature at which $J_s$ vanishes starting from low temperatures. For the case $f=0$, we used instead the Nelson-Kosterlitz universal relation~\cite{nelsonUniversalJumpSuperfluid1977}. \label{fig6}}
\end{figure*}

\section{Discussion}

By means of Monte Carlo simulations on the uniformly frustrated $XY$ model, we have shown that the presence of disorder, mimicked via random couplings, modifies the ordered vortex lattice of the ground state. 
For strong enough disorder, the energy of the system is no longer minimized by an ordered vortex-lattice structure, but rather by a disordered structure in which the vortex cores are located where the local stiffness of the superfluid is lower so as to reduce the energy cost associated with a phase twist.

At the same time, we have shown that in the presence of a very weak pinning as in the case with no disorder, where the only pinning potential is due to the presence of the underlying  square grid, the increase of the applied magnetic field does not reduce the superconducting critical temperature, but in most cases it contributes to its increase, in stark contrast with the  experimental observations~\cite{yazdaniObservationKosterlitzThoulesstypeMelting1993, blatterVorticesHightemperatureSuperconductors1994, feigelmanVorticesHightemperatureSuperconductors1994, chenTwodimensionalVorticesSuperconductors2007, guillamonDirectObservationMelting2009, misraMeasurementsMagneticFieldTunedConductivity2013, benyaminiFragilityDissipationlessState2019, royMeltingVortexLattice2019}.
The presence of disorder not only restores the experimentally observed dependence between $T_c$ and $f$, but also acts by making the ground-state vortex lattice more robust against thermal fluctuations. Comparing two different levels of disorder, we have indeed shown that the superconducting critical temperature is suppressed much less with respect to the zero-field case in the strong disorder regime than in the weak disorder regime. 

In conclusion, despite separately both disorder and magnetic field act on the SC thin film by suppressing the superfluid stiffness, when a transverse magnetic field is applied to the system, the presence of disorder help to prevent the destruction of the SC state by thermal fluctuations.


\bibliography{maccari}

\begin{thebibliography}{43}%
\makeatletter
\providecommand \@ifxundefined [1]{%
 \@ifx{#1\undefined}
}%
\providecommand \@ifnum [1]{%
 \ifnum #1\expandafter \@firstoftwo
 \else \expandafter \@secondoftwo
 \fi
}%
\providecommand \@ifx [1]{%
 \ifx #1\expandafter \@firstoftwo
 \else \expandafter \@secondoftwo
 \fi
}%
\providecommand \natexlab [1]{#1}%
\providecommand \enquote  [1]{``#1''}%
\providecommand \bibnamefont  [1]{#1}%
\providecommand \bibfnamefont [1]{#1}%
\providecommand \citenamefont [1]{#1}%
\providecommand \href@noop [0]{\@secondoftwo}%
\providecommand \href [0]{\begingroup \@sanitize@url \@href}%
\providecommand \@href[1]{\@@startlink{#1}\@@href}%
\providecommand \@@href[1]{\endgroup#1\@@endlink}%
\providecommand \@sanitize@url [0]{\catcode `\\12\catcode `\$12\catcode
  `\&12\catcode `\#12\catcode `\^12\catcode `\_12\catcode `\%12\relax}%
\providecommand \@@startlink[1]{}%
\providecommand \@@endlink[0]{}%
\providecommand \url  [0]{\begingroup\@sanitize@url \@url }%
\providecommand \@url [1]{\endgroup\@href {#1}{\urlprefix }}%
\providecommand \urlprefix  [0]{URL }%
\providecommand \Eprint [0]{\href }%
\providecommand \doibase [0]{http://dx.doi.org/}%
\providecommand \selectlanguage [0]{\@gobble}%
\providecommand \bibinfo  [0]{\@secondoftwo}%
\providecommand \bibfield  [0]{\@secondoftwo}%
\providecommand \translation [1]{[#1]}%
\providecommand \BibitemOpen [0]{}%
\providecommand \bibitemStop [0]{}%
\providecommand \bibitemNoStop [0]{.\EOS\space}%
\providecommand \EOS [0]{\spacefactor3000\relax}%
\providecommand \BibitemShut  [1]{\csname bibitem#1\endcsname}%
\let\auto@bib@innerbib\@empty
\bibitem [{\citenamefont {Fisher}\ \emph {et~al.}(1991)\citenamefont {Fisher},
  \citenamefont {Fisher},\ and\ \citenamefont
  {Huse}}]{fisherThermalFluctuationsQuenched1991}%
  \BibitemOpen
  \bibfield  {author} {\bibinfo {author} {\bibfnamefont {D.~S.}\ \bibnamefont
  {Fisher}}, \bibinfo {author} {\bibfnamefont {M.~P.~A.}\ \bibnamefont
  {Fisher}}, \ and\ \bibinfo {author} {\bibfnamefont {D.~A.}\ \bibnamefont
  {Huse}},\ }\href {\doibase 10.1103/PhysRevB.43.130} {\bibfield  {journal}
  {\bibinfo  {journal} {Physical Review B}\ }\textbf {\bibinfo {volume} {43}},\
  \bibinfo {pages} {130} (\bibinfo {year} {1991})},\ \bibinfo {note}
  {publisher: American Physical Society}\BibitemShut {NoStop}%
\bibitem [{\citenamefont {Blatter}\ \emph {et~al.}(1994)\citenamefont
  {Blatter}, \citenamefont {Feigel'man}, \citenamefont {Geshkenbein},
  \citenamefont {Larkin},\ and\ \citenamefont
  {Vinokur}}]{blatterVorticesHightemperatureSuperconductors1994}%
  \BibitemOpen
  \bibfield  {author} {\bibinfo {author} {\bibfnamefont {G.}~\bibnamefont
  {Blatter}}, \bibinfo {author} {\bibfnamefont {M.~V.}\ \bibnamefont
  {Feigel'man}}, \bibinfo {author} {\bibfnamefont {V.~B.}\ \bibnamefont
  {Geshkenbein}}, \bibinfo {author} {\bibfnamefont {A.~I.}\ \bibnamefont
  {Larkin}}, \ and\ \bibinfo {author} {\bibfnamefont {V.~M.}\ \bibnamefont
  {Vinokur}},\ }\href {\doibase 10.1103/RevModPhys.66.1125} {\bibfield
  {journal} {\bibinfo  {journal} {Reviews of Modern Physics}\ }\textbf
  {\bibinfo {volume} {66}},\ \bibinfo {pages} {1125} (\bibinfo {year}
  {1994})},\ \bibinfo {note} {publisher: American Physical Society}\BibitemShut
  {NoStop}%
\bibitem [{\citenamefont {Klein}\ \emph {et~al.}(2001)\citenamefont {Klein},
  \citenamefont {Joumard}, \citenamefont {Blanchard}, \citenamefont {Marcus},
  \citenamefont {Cubitt}, \citenamefont {Giamarchi},\ and\ \citenamefont
  {Le~Doussal}}]{kleinBraggGlassPhase2001}%
  \BibitemOpen
  \bibfield  {author} {\bibinfo {author} {\bibfnamefont {T.}~\bibnamefont
  {Klein}}, \bibinfo {author} {\bibfnamefont {I.}~\bibnamefont {Joumard}},
  \bibinfo {author} {\bibfnamefont {S.}~\bibnamefont {Blanchard}}, \bibinfo
  {author} {\bibfnamefont {J.}~\bibnamefont {Marcus}}, \bibinfo {author}
  {\bibfnamefont {R.}~\bibnamefont {Cubitt}}, \bibinfo {author} {\bibfnamefont
  {T.}~\bibnamefont {Giamarchi}}, \ and\ \bibinfo {author} {\bibfnamefont
  {P.}~\bibnamefont {Le~Doussal}},\ }\href {\doibase 10.1038/35096534}
  {\bibfield  {journal} {\bibinfo  {journal} {Nature}\ }\textbf {\bibinfo
  {volume} {413}},\ \bibinfo {pages} {404} (\bibinfo {year} {2001})},\ \bibinfo
  {note} {bandiera\_abtest: a Cg\_type: Nature Research Journals Number: 6854
  Primary\_atype: Research Publisher: Nature Publishing Group}\BibitemShut
  {NoStop}%
\bibitem [{\citenamefont {Guillamón}\ \emph {et~al.}(2009)\citenamefont
  {Guillamón}, \citenamefont {Suderow}, \citenamefont {Fernández-Pacheco},
  \citenamefont {Sesé}, \citenamefont {Córdoba}, \citenamefont {De~Teresa},
  \citenamefont {Ibarra},\ and\ \citenamefont
  {Vieira}}]{guillamonDirectObservationMelting2009}%
  \BibitemOpen
  \bibfield  {author} {\bibinfo {author} {\bibfnamefont {I.}~\bibnamefont
  {Guillamón}}, \bibinfo {author} {\bibfnamefont {H.}~\bibnamefont {Suderow}},
  \bibinfo {author} {\bibfnamefont {A.}~\bibnamefont {Fernández-Pacheco}},
  \bibinfo {author} {\bibfnamefont {J.}~\bibnamefont {Sesé}}, \bibinfo
  {author} {\bibfnamefont {R.}~\bibnamefont {Córdoba}}, \bibinfo {author}
  {\bibfnamefont {J.~M.}\ \bibnamefont {De~Teresa}}, \bibinfo {author}
  {\bibfnamefont {M.~R.}\ \bibnamefont {Ibarra}}, \ and\ \bibinfo {author}
  {\bibfnamefont {S.}~\bibnamefont {Vieira}},\ }\href {\doibase
  10.1038/nphys1368} {\bibfield  {journal} {\bibinfo  {journal} {Nature
  Physics}\ }\textbf {\bibinfo {volume} {5}},\ \bibinfo {pages} {651} (\bibinfo
  {year} {2009})},\ \bibinfo {note} {number: 9 Publisher: Nature Publishing
  Group}\BibitemShut {NoStop}%
\bibitem [{\citenamefont {Roy}\ \emph {et~al.}(2019)\citenamefont {Roy},
  \citenamefont {Dutta}, \citenamefont {Roy~Choudhury}, \citenamefont
  {Basistha}, \citenamefont {Maccari}, \citenamefont {Mandal}, \citenamefont
  {Jesudasan}, \citenamefont {Bagwe}, \citenamefont {Castellani}, \citenamefont
  {Benfatto},\ and\ \citenamefont
  {Raychaudhuri}}]{royMeltingVortexLattice2019}%
  \BibitemOpen
  \bibfield  {author} {\bibinfo {author} {\bibfnamefont {I.}~\bibnamefont
  {Roy}}, \bibinfo {author} {\bibfnamefont {S.}~\bibnamefont {Dutta}}, \bibinfo
  {author} {\bibfnamefont {A.~N.}\ \bibnamefont {Roy~Choudhury}}, \bibinfo
  {author} {\bibfnamefont {S.}~\bibnamefont {Basistha}}, \bibinfo {author}
  {\bibfnamefont {I.}~\bibnamefont {Maccari}}, \bibinfo {author} {\bibfnamefont
  {S.}~\bibnamefont {Mandal}}, \bibinfo {author} {\bibfnamefont
  {J.}~\bibnamefont {Jesudasan}}, \bibinfo {author} {\bibfnamefont
  {V.}~\bibnamefont {Bagwe}}, \bibinfo {author} {\bibfnamefont
  {C.}~\bibnamefont {Castellani}}, \bibinfo {author} {\bibfnamefont
  {L.}~\bibnamefont {Benfatto}}, \ and\ \bibinfo {author} {\bibfnamefont
  {P.}~\bibnamefont {Raychaudhuri}},\ }\href {\doibase
  10.1103/PhysRevLett.122.047001} {\bibfield  {journal} {\bibinfo  {journal}
  {Physical Review Letters}\ }\textbf {\bibinfo {volume} {122}},\ \bibinfo
  {pages} {047001} (\bibinfo {year} {2019})}\BibitemShut {NoStop}%
\bibitem [{\citenamefont
  {Berezinsky}(1972)}]{berezinskyDestructionLongrangeOrder1972}%
  \BibitemOpen
  \bibfield  {author} {\bibinfo {author} {\bibfnamefont {V.~L.}\ \bibnamefont
  {Berezinsky}},\ }\href@noop {} {\bibfield  {journal} {\bibinfo  {journal}
  {Sov. Phys. JETP}\ }\textbf {\bibinfo {volume} {34}},\ \bibinfo {pages} {610}
  (\bibinfo {year} {1972})}\BibitemShut {NoStop}%
\bibitem [{\citenamefont {Kosterlitz}\ and\ \citenamefont
  {Thouless}(1973)}]{kosterlitzOrderingMetastabilityPhase1973}%
  \BibitemOpen
  \bibfield  {author} {\bibinfo {author} {\bibfnamefont {J.~M.}\ \bibnamefont
  {Kosterlitz}}\ and\ \bibinfo {author} {\bibfnamefont {D.~J.}\ \bibnamefont
  {Thouless}},\ }\href {\doibase 10.1088/0022-3719/6/7/010} {\bibfield
  {journal} {\bibinfo  {journal} {Journal of Physics C: Solid State Physics}\
  }\textbf {\bibinfo {volume} {6}},\ \bibinfo {pages} {1181} (\bibinfo {year}
  {1973})}\BibitemShut {NoStop}%
\bibitem [{\citenamefont
  {Kosterlitz}(1974)}]{kosterlitzCriticalPropertiesTwodimensional1974}%
  \BibitemOpen
  \bibfield  {author} {\bibinfo {author} {\bibfnamefont {J.~M.}\ \bibnamefont
  {Kosterlitz}},\ }\href {\doibase 10.1088/0022-3719/7/6/005} {\bibfield
  {journal} {\bibinfo  {journal} {Journal of Physics C: Solid State Physics}\
  }\textbf {\bibinfo {volume} {7}},\ \bibinfo {pages} {1046} (\bibinfo {year}
  {1974})},\ \bibinfo {note} {publisher: IOP Publishing}\BibitemShut {NoStop}%
\bibitem [{\citenamefont {Halperin}\ and\ \citenamefont
  {Nelson}(1978)}]{halperinTheoryTwoDimensionalMelting1978}%
  \BibitemOpen
  \bibfield  {author} {\bibinfo {author} {\bibfnamefont {B.~I.}\ \bibnamefont
  {Halperin}}\ and\ \bibinfo {author} {\bibfnamefont {D.~R.}\ \bibnamefont
  {Nelson}},\ }\href {\doibase 10.1103/PhysRevLett.41.121} {\bibfield
  {journal} {\bibinfo  {journal} {Physical Review Letters}\ }\textbf {\bibinfo
  {volume} {41}},\ \bibinfo {pages} {121} (\bibinfo {year} {1978})},\ \bibinfo
  {note} {publisher: American Physical Society}\BibitemShut {NoStop}%
\bibitem [{\citenamefont {Nelson}\ and\ \citenamefont
  {Halperin}(1979)}]{nelsonDislocationmediatedMeltingTwo1979}%
  \BibitemOpen
  \bibfield  {author} {\bibinfo {author} {\bibfnamefont {D.~R.}\ \bibnamefont
  {Nelson}}\ and\ \bibinfo {author} {\bibfnamefont {B.~I.}\ \bibnamefont
  {Halperin}},\ }\href {\doibase 10.1103/PhysRevB.19.2457} {\bibfield
  {journal} {\bibinfo  {journal} {Physical Review B}\ }\textbf {\bibinfo
  {volume} {19}},\ \bibinfo {pages} {2457} (\bibinfo {year} {1979})},\ \bibinfo
  {note} {publisher: American Physical Society}\BibitemShut {NoStop}%
\bibitem [{\citenamefont {Young}(1979)}]{youngMeltingVectorCoulomb1979}%
  \BibitemOpen
  \bibfield  {author} {\bibinfo {author} {\bibfnamefont {A.~P.}\ \bibnamefont
  {Young}},\ }\href {\doibase 10.1103/PhysRevB.19.1855} {\bibfield  {journal}
  {\bibinfo  {journal} {Physical Review B}\ }\textbf {\bibinfo {volume} {19}},\
  \bibinfo {pages} {1855} (\bibinfo {year} {1979})},\ \bibinfo {note}
  {publisher: American Physical Society}\BibitemShut {NoStop}%
\bibitem [{\citenamefont {Giamarchi}\ and\ \citenamefont
  {Le~Doussal}(1995)}]{giamarchiElasticTheoryFlux1995}%
  \BibitemOpen
  \bibfield  {author} {\bibinfo {author} {\bibfnamefont {T.}~\bibnamefont
  {Giamarchi}}\ and\ \bibinfo {author} {\bibfnamefont {P.}~\bibnamefont
  {Le~Doussal}},\ }\href {\doibase 10.1103/PhysRevB.52.1242} {\bibfield
  {journal} {\bibinfo  {journal} {Physical Review B}\ }\textbf {\bibinfo
  {volume} {52}},\ \bibinfo {pages} {1242} (\bibinfo {year} {1995})},\ \bibinfo
  {note} {publisher: American Physical Society}\BibitemShut {NoStop}%
\bibitem [{\citenamefont {Le~Doussal}\ and\ \citenamefont
  {Giamarchi}(2000)}]{ledoussalDislocationsBraggGlasses2000}%
  \BibitemOpen
  \bibfield  {author} {\bibinfo {author} {\bibfnamefont {P.}~\bibnamefont
  {Le~Doussal}}\ and\ \bibinfo {author} {\bibfnamefont {T.}~\bibnamefont
  {Giamarchi}},\ }\href {\doibase 10.1016/S0921-4534(00)00005-8} {\bibfield
  {journal} {\bibinfo  {journal} {Physica C: Superconductivity}\ }\textbf
  {\bibinfo {volume} {331}},\ \bibinfo {pages} {233} (\bibinfo {year}
  {2000})}\BibitemShut {NoStop}%
\bibitem [{\citenamefont {Ganguly}\ \emph {et~al.}(2017)\citenamefont
  {Ganguly}, \citenamefont {Roy}, \citenamefont {Banerjee}, \citenamefont
  {Singh}, \citenamefont {Ghosal},\ and\ \citenamefont
  {Raychaudhuri}}]{gangulyMagneticFieldInduced2017}%
  \BibitemOpen
  \bibfield  {author} {\bibinfo {author} {\bibfnamefont {R.}~\bibnamefont
  {Ganguly}}, \bibinfo {author} {\bibfnamefont {I.}~\bibnamefont {Roy}},
  \bibinfo {author} {\bibfnamefont {A.}~\bibnamefont {Banerjee}}, \bibinfo
  {author} {\bibfnamefont {H.}~\bibnamefont {Singh}}, \bibinfo {author}
  {\bibfnamefont {A.}~\bibnamefont {Ghosal}}, \ and\ \bibinfo {author}
  {\bibfnamefont {P.}~\bibnamefont {Raychaudhuri}},\ }\href {\doibase
  10.1103/PhysRevB.96.054509} {\bibfield  {journal} {\bibinfo  {journal}
  {Physical Review B}\ }\textbf {\bibinfo {volume} {96}},\ \bibinfo {pages}
  {054509} (\bibinfo {year} {2017})},\ \bibinfo {note} {publisher: American
  Physical Society}\BibitemShut {NoStop}%
\bibitem [{\citenamefont {Barabash}\ \emph {et~al.}(2000)\citenamefont
  {Barabash}, \citenamefont {Stroud},\ and\ \citenamefont
  {Hwang}}]{barabashConductivityDueClassical2000}%
  \BibitemOpen
  \bibfield  {author} {\bibinfo {author} {\bibfnamefont {S.}~\bibnamefont
  {Barabash}}, \bibinfo {author} {\bibfnamefont {D.}~\bibnamefont {Stroud}}, \
  and\ \bibinfo {author} {\bibfnamefont {I.-J.}\ \bibnamefont {Hwang}},\ }\href
  {\doibase 10.1103/PhysRevB.61.R14924} {\bibfield  {journal} {\bibinfo
  {journal} {Physical Review B}\ }\textbf {\bibinfo {volume} {61}},\ \bibinfo
  {pages} {R14924} (\bibinfo {year} {2000})},\ \bibinfo {note} {publisher:
  American Physical Society}\BibitemShut {NoStop}%
\bibitem [{\citenamefont {Wysin}\ \emph {et~al.}(2005)\citenamefont {Wysin},
  \citenamefont {Pereira}, \citenamefont {Marques}, \citenamefont {Leonel},\
  and\ \citenamefont
  {Coura}}]{wysinExtinctionBerezinskiiKosterlitzThoulessPhase2005}%
  \BibitemOpen
  \bibfield  {author} {\bibinfo {author} {\bibfnamefont {G.~M.}\ \bibnamefont
  {Wysin}}, \bibinfo {author} {\bibfnamefont {A.~R.}\ \bibnamefont {Pereira}},
  \bibinfo {author} {\bibfnamefont {I.~A.}\ \bibnamefont {Marques}}, \bibinfo
  {author} {\bibfnamefont {S.~A.}\ \bibnamefont {Leonel}}, \ and\ \bibinfo
  {author} {\bibfnamefont {P.~Z.}\ \bibnamefont {Coura}},\ }\href {\doibase
  10.1103/PhysRevB.72.094418} {\bibfield  {journal} {\bibinfo  {journal}
  {Physical Review B}\ }\textbf {\bibinfo {volume} {72}},\ \bibinfo {pages}
  {094418} (\bibinfo {year} {2005})},\ \bibinfo {note} {publisher: American
  Physical Society}\BibitemShut {NoStop}%
\bibitem [{\citenamefont {Erez}\ and\ \citenamefont
  {Meir}(2013)}]{erezEffectAmplitudeFluctuations2013}%
  \BibitemOpen
  \bibfield  {author} {\bibinfo {author} {\bibfnamefont {A.}~\bibnamefont
  {Erez}}\ and\ \bibinfo {author} {\bibfnamefont {Y.}~\bibnamefont {Meir}},\
  }\href {\doibase 10.1103/PhysRevB.88.184510} {\bibfield  {journal} {\bibinfo
  {journal} {Physical Review B}\ }\textbf {\bibinfo {volume} {88}},\ \bibinfo
  {pages} {184510} (\bibinfo {year} {2013})},\ \bibinfo {note} {publisher:
  American Physical Society}\BibitemShut {NoStop}%
\bibitem [{\citenamefont {Costa}\ \emph {et~al.}(2014)\citenamefont {Costa},
  \citenamefont {Lima}, \citenamefont {Coura}, \citenamefont {Leonel},\ and\
  \citenamefont {Lima}}]{costaKosterlitzThoulessTransitionDiluted2014}%
  \BibitemOpen
  \bibfield  {author} {\bibinfo {author} {\bibfnamefont {B.~V.}\ \bibnamefont
  {Costa}}, \bibinfo {author} {\bibfnamefont {L.~S.}\ \bibnamefont {Lima}},
  \bibinfo {author} {\bibfnamefont {P.~Z.}\ \bibnamefont {Coura}}, \bibinfo
  {author} {\bibfnamefont {S.~A.}\ \bibnamefont {Leonel}}, \ and\ \bibinfo
  {author} {\bibfnamefont {A.~B.}\ \bibnamefont {Lima}},\ }\href {\doibase
  10.1088/1742-6596/487/1/012008} {\bibfield  {journal} {\bibinfo  {journal}
  {Journal of Physics: Conference Series}\ }\textbf {\bibinfo {volume} {487}},\
  \bibinfo {pages} {012008} (\bibinfo {year} {2014})},\ \bibinfo {note}
  {publisher: IOP Publishing}\BibitemShut {NoStop}%
\bibitem [{\citenamefont {Maccari}\ \emph {et~al.}(2016)\citenamefont
  {Maccari}, \citenamefont {Maiorano}, \citenamefont {Marinari},\ and\
  \citenamefont {Ruiz-Lorenzo}}]{maccariNumericalStudyPlanar2016}%
  \BibitemOpen
  \bibfield  {author} {\bibinfo {author} {\bibfnamefont {I.}~\bibnamefont
  {Maccari}}, \bibinfo {author} {\bibfnamefont {A.}~\bibnamefont {Maiorano}},
  \bibinfo {author} {\bibfnamefont {E.}~\bibnamefont {Marinari}}, \ and\
  \bibinfo {author} {\bibfnamefont {J.~J.}\ \bibnamefont {Ruiz-Lorenzo}},\
  }\href {\doibase 10.1140/epjb/e2016-70171-x} {\bibfield  {journal} {\bibinfo
  {journal} {The European Physical Journal B}\ }\textbf {\bibinfo {volume}
  {89}},\ \bibinfo {pages} {127} (\bibinfo {year} {2016})}\BibitemShut
  {NoStop}%
\bibitem [{\citenamefont {Kumar}\ \emph {et~al.}(2017)\citenamefont {Kumar},
  \citenamefont {Chatterjee}, \citenamefont {Paul},\ and\ \citenamefont
  {Puri}}]{kumarOrderingKineticsRandombond2017}%
  \BibitemOpen
  \bibfield  {author} {\bibinfo {author} {\bibfnamefont {M.}~\bibnamefont
  {Kumar}}, \bibinfo {author} {\bibfnamefont {S.}~\bibnamefont {Chatterjee}},
  \bibinfo {author} {\bibfnamefont {R.}~\bibnamefont {Paul}}, \ and\ \bibinfo
  {author} {\bibfnamefont {S.}~\bibnamefont {Puri}},\ }\href {\doibase
  10.1103/PhysRevE.96.042127} {\bibfield  {journal} {\bibinfo  {journal}
  {Physical Review E}\ }\textbf {\bibinfo {volume} {96}},\ \bibinfo {pages}
  {042127} (\bibinfo {year} {2017})},\ \bibinfo {note} {publisher: American
  Physical Society}\BibitemShut {NoStop}%
\bibitem [{\citenamefont {Maccari}\ \emph {et~al.}(2017)\citenamefont
  {Maccari}, \citenamefont {Benfatto},\ and\ \citenamefont
  {Castellani}}]{maccariBroadeningBerezinskiiKosterlitzThoulessTransition2017}%
  \BibitemOpen
  \bibfield  {author} {\bibinfo {author} {\bibfnamefont {I.}~\bibnamefont
  {Maccari}}, \bibinfo {author} {\bibfnamefont {L.}~\bibnamefont {Benfatto}}, \
  and\ \bibinfo {author} {\bibfnamefont {C.}~\bibnamefont {Castellani}},\
  }\href {\doibase 10.1103/PhysRevB.96.060508} {\bibfield  {journal} {\bibinfo
  {journal} {Physical Review B}\ }\textbf {\bibinfo {volume} {96}},\ \bibinfo
  {pages} {060508} (\bibinfo {year} {2017})}\BibitemShut {NoStop}%
\bibitem [{\citenamefont {Maccari}\ \emph {et~al.}(2019)\citenamefont
  {Maccari}, \citenamefont {Benfatto},\ and\ \citenamefont
  {Castellani}}]{maccariDisorderedXYModel2019}%
  \BibitemOpen
  \bibfield  {author} {\bibinfo {author} {\bibfnamefont {I.}~\bibnamefont
  {Maccari}}, \bibinfo {author} {\bibfnamefont {L.}~\bibnamefont {Benfatto}}, \
  and\ \bibinfo {author} {\bibfnamefont {C.}~\bibnamefont {Castellani}},\
  }\href {\doibase 10.1103/PhysRevB.99.104509} {\bibfield  {journal} {\bibinfo
  {journal} {Physical Review B}\ }\textbf {\bibinfo {volume} {99}},\ \bibinfo
  {pages} {104509} (\bibinfo {year} {2019})}\BibitemShut {NoStop}%
\bibitem [{\citenamefont {Teitel}\ and\ \citenamefont
  {Jayaprakash}(1983)}]{teitelPhaseTranstionsFrustrated1983}%
  \BibitemOpen
  \bibfield  {author} {\bibinfo {author} {\bibfnamefont {S.}~\bibnamefont
  {Teitel}}\ and\ \bibinfo {author} {\bibfnamefont {C.}~\bibnamefont
  {Jayaprakash}},\ }\href {\doibase 10.1103/PhysRevB.27.598} {\bibfield
  {journal} {\bibinfo  {journal} {Physical Review B}\ }\textbf {\bibinfo
  {volume} {27}},\ \bibinfo {pages} {598} (\bibinfo {year} {1983})},\ \bibinfo
  {note} {publisher: American Physical Society}\BibitemShut {NoStop}%
\bibitem [{\citenamefont {Franz}\ and\ \citenamefont
  {Teitel}(1994)}]{franzVortexLatticeMelting1994}%
  \BibitemOpen
  \bibfield  {author} {\bibinfo {author} {\bibfnamefont {M.}~\bibnamefont
  {Franz}}\ and\ \bibinfo {author} {\bibfnamefont {S.}~\bibnamefont {Teitel}},\
  }\href {\doibase 10.1103/PhysRevLett.73.480} {\bibfield  {journal} {\bibinfo
  {journal} {Physical Review Letters}\ }\textbf {\bibinfo {volume} {73}},\
  \bibinfo {pages} {480} (\bibinfo {year} {1994})},\ \bibinfo {note}
  {publisher: American Physical Society}\BibitemShut {NoStop}%
\bibitem [{\citenamefont {Franz}\ and\ \citenamefont
  {Teitel}(1995)}]{franzVortexlatticeMeltingTwodimensional1995}%
  \BibitemOpen
  \bibfield  {author} {\bibinfo {author} {\bibfnamefont {M.}~\bibnamefont
  {Franz}}\ and\ \bibinfo {author} {\bibfnamefont {S.}~\bibnamefont {Teitel}},\
  }\href {\doibase 10.1103/PhysRevB.51.6551} {\bibfield  {journal} {\bibinfo
  {journal} {Physical Review B}\ }\textbf {\bibinfo {volume} {51}},\ \bibinfo
  {pages} {6551} (\bibinfo {year} {1995})}\BibitemShut {NoStop}%
\bibitem [{\citenamefont {Hattel}\ and\ \citenamefont
  {Wheatley}(1995)}]{hattelFluxlatticeMeltingDepinning1995}%
  \BibitemOpen
  \bibfield  {author} {\bibinfo {author} {\bibfnamefont {S.~A.}\ \bibnamefont
  {Hattel}}\ and\ \bibinfo {author} {\bibfnamefont {J.~M.}\ \bibnamefont
  {Wheatley}},\ }\href {\doibase 10.1103/PhysRevB.51.11951} {\bibfield
  {journal} {\bibinfo  {journal} {Physical Review B}\ }\textbf {\bibinfo
  {volume} {51}},\ \bibinfo {pages} {11951} (\bibinfo {year} {1995})},\
  \bibinfo {note} {publisher: American Physical Society}\BibitemShut {NoStop}%
\bibitem [{\citenamefont {Tanaka}\ and\ \citenamefont
  {Hu}(2001)}]{tanakaNumericalStudyFlux2001}%
  \BibitemOpen
  \bibfield  {author} {\bibinfo {author} {\bibfnamefont {A.}~\bibnamefont
  {Tanaka}}\ and\ \bibinfo {author} {\bibfnamefont {X.}~\bibnamefont {Hu}},\
  }\href {\doibase 10.1016/S0921-4534(01)00269-6} {\bibfield  {journal}
  {\bibinfo  {journal} {Physica C: Superconductivity}\ }\textbf {\bibinfo
  {volume} {357-360}},\ \bibinfo {pages} {438} (\bibinfo {year}
  {2001})}\BibitemShut {NoStop}%
\bibitem [{\citenamefont {Hasenbusch}\ \emph {et~al.}(2005)\citenamefont
  {Hasenbusch}, \citenamefont {Pelissetto},\ and\ \citenamefont
  {Vicari}}]{hasenbuschMulticriticalBehaviourFully2005}%
  \BibitemOpen
  \bibfield  {author} {\bibinfo {author} {\bibfnamefont {M.}~\bibnamefont
  {Hasenbusch}}, \bibinfo {author} {\bibfnamefont {A.}~\bibnamefont
  {Pelissetto}}, \ and\ \bibinfo {author} {\bibfnamefont {E.}~\bibnamefont
  {Vicari}},\ }\href {\doibase 10.1088/1742-5468/2005/12/P12002} {\bibfield
  {journal} {\bibinfo  {journal} {Journal of Statistical Mechanics: Theory and
  Experiment}\ }\textbf {\bibinfo {volume} {2005}},\ \bibinfo {pages} {P12002}
  (\bibinfo {year} {2005})}\BibitemShut {NoStop}%
\bibitem [{\citenamefont {Alba}\ \emph {et~al.}(2008)\citenamefont {Alba},
  \citenamefont {Pelissetto},\ and\ \citenamefont
  {Vicari}}]{albaUniformlyFrustratedTwodimensionalXYmodel2008}%
  \BibitemOpen
  \bibfield  {author} {\bibinfo {author} {\bibfnamefont {V.}~\bibnamefont
  {Alba}}, \bibinfo {author} {\bibfnamefont {A.}~\bibnamefont {Pelissetto}}, \
  and\ \bibinfo {author} {\bibfnamefont {E.}~\bibnamefont {Vicari}},\ }\href
  {\doibase 10.1088/1751-8113/41/17/175001} {\bibfield  {journal} {\bibinfo
  {journal} {Journal of Physics A: Mathematical and Theoretical}\ }\textbf
  {\bibinfo {volume} {41}},\ \bibinfo {pages} {175001} (\bibinfo {year}
  {2008})},\ \bibinfo {note} {publisher: IOP Publishing}\BibitemShut {NoStop}%
\bibitem [{\citenamefont
  {Teitel}(2013)}]{teitelTwoDimensionalFullyFrustrated2013}%
  \BibitemOpen
  \bibfield  {author} {\bibinfo {author} {\bibfnamefont {S.}~\bibnamefont
  {Teitel}},\ }in\ \href {\doibase 10.1142/9789814417648_0006} {\emph {\bibinfo
  {booktitle} {40 {Years} of {Berezinskii}–{Kosterlitz}–{Thouless}
  {Theory}}}}\ (\bibinfo  {publisher} {WORLD SCIENTIFIC},\ \bibinfo {year}
  {2013})\ pp.\ \bibinfo {pages} {201--235}\BibitemShut {NoStop}%
\bibitem [{\citenamefont {Alba}\ \emph {et~al.}(2009)\citenamefont {Alba},
  \citenamefont {Pelissetto},\ and\ \citenamefont
  {Vicari}}]{albaQuasilongrangeOrder2D2009}%
  \BibitemOpen
  \bibfield  {author} {\bibinfo {author} {\bibfnamefont {V.}~\bibnamefont
  {Alba}}, \bibinfo {author} {\bibfnamefont {A.}~\bibnamefont {Pelissetto}}, \
  and\ \bibinfo {author} {\bibfnamefont {E.}~\bibnamefont {Vicari}},\ }\href
  {\doibase 10.1088/1751-8113/42/29/295001} {\bibfield  {journal} {\bibinfo
  {journal} {Journal of Physics A: Mathematical and Theoretical}\ }\textbf
  {\bibinfo {volume} {42}},\ \bibinfo {pages} {295001} (\bibinfo {year}
  {2009})}\BibitemShut {NoStop}%
\bibitem [{\citenamefont {Alba}\ \emph {et~al.}(2010)\citenamefont {Alba},
  \citenamefont {Pelissetto},\ and\ \citenamefont
  {Vicari}}]{albaMagneticGlassyTransitions2010}%
  \BibitemOpen
  \bibfield  {author} {\bibinfo {author} {\bibfnamefont {V.}~\bibnamefont
  {Alba}}, \bibinfo {author} {\bibfnamefont {A.}~\bibnamefont {Pelissetto}}, \
  and\ \bibinfo {author} {\bibfnamefont {E.}~\bibnamefont {Vicari}},\ }\href
  {\doibase 10.1088/1742-5468/2010/03/P03006} {\bibfield  {journal} {\bibinfo
  {journal} {Journal of Statistical Mechanics: Theory and Experiment}\ }\textbf
  {\bibinfo {volume} {2010}},\ \bibinfo {pages} {P03006} (\bibinfo {year}
  {2010})},\ \bibinfo {note} {publisher: IOP Publishing}\BibitemShut {NoStop}%
\bibitem [{\citenamefont {Yazdani}\ \emph {et~al.}(1993)\citenamefont
  {Yazdani}, \citenamefont {White}, \citenamefont {Hahn}, \citenamefont
  {Gabay}, \citenamefont {Beasley},\ and\ \citenamefont
  {Kapitulnik}}]{yazdaniObservationKosterlitzThoulesstypeMelting1993}%
  \BibitemOpen
  \bibfield  {author} {\bibinfo {author} {\bibfnamefont {A.}~\bibnamefont
  {Yazdani}}, \bibinfo {author} {\bibfnamefont {W.~R.}\ \bibnamefont {White}},
  \bibinfo {author} {\bibfnamefont {M.~R.}\ \bibnamefont {Hahn}}, \bibinfo
  {author} {\bibfnamefont {M.}~\bibnamefont {Gabay}}, \bibinfo {author}
  {\bibfnamefont {M.~R.}\ \bibnamefont {Beasley}}, \ and\ \bibinfo {author}
  {\bibfnamefont {A.}~\bibnamefont {Kapitulnik}},\ }\href {\doibase
  10.1103/PhysRevLett.70.505} {\bibfield  {journal} {\bibinfo  {journal}
  {Physical Review Letters}\ }\textbf {\bibinfo {volume} {70}},\ \bibinfo
  {pages} {505} (\bibinfo {year} {1993})},\ \bibinfo {note} {publisher:
  American Physical Society}\BibitemShut {NoStop}%
\bibitem [{\citenamefont {Feigel'man}\ \emph {et~al.}(1994)\citenamefont
  {Feigel'man}, \citenamefont {Geshkenbein}, \citenamefont {Larkin},
  \citenamefont {Vinokur},\ and\ \citenamefont
  {Blatter}}]{feigelmanVorticesHightemperatureSuperconductors1994}%
  \BibitemOpen
  \bibfield  {author} {\bibinfo {author} {\bibfnamefont {M.~V.}\ \bibnamefont
  {Feigel'man}}, \bibinfo {author} {\bibfnamefont {V.~B.}\ \bibnamefont
  {Geshkenbein}}, \bibinfo {author} {\bibfnamefont {A.~I.}\ \bibnamefont
  {Larkin}}, \bibinfo {author} {\bibfnamefont {V.~M.}\ \bibnamefont {Vinokur}},
  \ and\ \bibinfo {author} {\bibfnamefont {G.}~\bibnamefont {Blatter}},\ }\href
  {\doibase 10.1103/RevModPhys.66.1125} {\bibfield  {journal} {\bibinfo
  {journal} {Reviews of Modern Physics}\ }\textbf {\bibinfo {volume} {66}},\
  \bibinfo {pages} {1125} (\bibinfo {year} {1994})},\ \bibinfo {note}
  {publisher: American Physical Society}\BibitemShut {NoStop}%
\bibitem [{\citenamefont {Chen}\ \emph {et~al.}(2007)\citenamefont {Chen},
  \citenamefont {Halperin}, \citenamefont {Guptasarma}, \citenamefont {Hinks},
  \citenamefont {Mitrović}, \citenamefont {Reyes},\ and\ \citenamefont
  {Kuhns}}]{chenTwodimensionalVorticesSuperconductors2007}%
  \BibitemOpen
  \bibfield  {author} {\bibinfo {author} {\bibfnamefont {B.}~\bibnamefont
  {Chen}}, \bibinfo {author} {\bibfnamefont {W.~P.}\ \bibnamefont {Halperin}},
  \bibinfo {author} {\bibfnamefont {P.}~\bibnamefont {Guptasarma}}, \bibinfo
  {author} {\bibfnamefont {D.~G.}\ \bibnamefont {Hinks}}, \bibinfo {author}
  {\bibfnamefont {V.~F.}\ \bibnamefont {Mitrović}}, \bibinfo {author}
  {\bibfnamefont {A.~P.}\ \bibnamefont {Reyes}}, \ and\ \bibinfo {author}
  {\bibfnamefont {P.~L.}\ \bibnamefont {Kuhns}},\ }\href {\doibase
  10.1038/nphys540} {\bibfield  {journal} {\bibinfo  {journal} {Nature
  Physics}\ }\textbf {\bibinfo {volume} {3}},\ \bibinfo {pages} {239} (\bibinfo
  {year} {2007})},\ \bibinfo {note} {bandiera\_abtest: a Cg\_type: Nature
  Research Journals Number: 4 Primary\_atype: Research Publisher: Nature
  Publishing Group}\BibitemShut {NoStop}%
\bibitem [{\citenamefont {Misra}\ \emph {et~al.}(2013)\citenamefont {Misra},
  \citenamefont {Urban}, \citenamefont {Kim}, \citenamefont {Sambandamurthy},\
  and\ \citenamefont
  {Yazdani}}]{misraMeasurementsMagneticFieldTunedConductivity2013}%
  \BibitemOpen
  \bibfield  {author} {\bibinfo {author} {\bibfnamefont {S.}~\bibnamefont
  {Misra}}, \bibinfo {author} {\bibfnamefont {L.}~\bibnamefont {Urban}},
  \bibinfo {author} {\bibfnamefont {M.}~\bibnamefont {Kim}}, \bibinfo {author}
  {\bibfnamefont {G.}~\bibnamefont {Sambandamurthy}}, \ and\ \bibinfo {author}
  {\bibfnamefont {A.}~\bibnamefont {Yazdani}},\ }\href {\doibase
  10.1103/PhysRevLett.110.037002} {\bibfield  {journal} {\bibinfo  {journal}
  {Physical Review Letters}\ }\textbf {\bibinfo {volume} {110}},\ \bibinfo
  {pages} {037002} (\bibinfo {year} {2013})},\ \bibinfo {note} {publisher:
  American Physical Society}\BibitemShut {NoStop}%
\bibitem [{\citenamefont {Benyamini}\ \emph {et~al.}(2019)\citenamefont
  {Benyamini}, \citenamefont {Telford}, \citenamefont {Kennes}, \citenamefont
  {Wang}, \citenamefont {Williams}, \citenamefont {Watanabe}, \citenamefont
  {Taniguchi}, \citenamefont {Shahar}, \citenamefont {Hone}, \citenamefont
  {Dean}, \citenamefont {Millis},\ and\ \citenamefont
  {Pasupathy}}]{benyaminiFragilityDissipationlessState2019}%
  \BibitemOpen
  \bibfield  {author} {\bibinfo {author} {\bibfnamefont {A.}~\bibnamefont
  {Benyamini}}, \bibinfo {author} {\bibfnamefont {E.~J.}\ \bibnamefont
  {Telford}}, \bibinfo {author} {\bibfnamefont {D.~M.}\ \bibnamefont {Kennes}},
  \bibinfo {author} {\bibfnamefont {D.}~\bibnamefont {Wang}}, \bibinfo {author}
  {\bibfnamefont {A.}~\bibnamefont {Williams}}, \bibinfo {author}
  {\bibfnamefont {K.}~\bibnamefont {Watanabe}}, \bibinfo {author}
  {\bibfnamefont {T.}~\bibnamefont {Taniguchi}}, \bibinfo {author}
  {\bibfnamefont {D.}~\bibnamefont {Shahar}}, \bibinfo {author} {\bibfnamefont
  {J.}~\bibnamefont {Hone}}, \bibinfo {author} {\bibfnamefont {C.~R.}\
  \bibnamefont {Dean}}, \bibinfo {author} {\bibfnamefont {A.~J.}\ \bibnamefont
  {Millis}}, \ and\ \bibinfo {author} {\bibfnamefont {A.~N.}\ \bibnamefont
  {Pasupathy}},\ }\href {\doibase 10.1038/s41567-019-0571-z} {\bibfield
  {journal} {\bibinfo  {journal} {Nature Physics}\ }\textbf {\bibinfo {volume}
  {15}},\ \bibinfo {pages} {947} (\bibinfo {year} {2019})},\ \bibinfo {note}
  {number: 9 Publisher: Nature Publishing Group}\BibitemShut {NoStop}%
\bibitem [{\citenamefont {Ma}\ and\ \citenamefont
  {Lee}(1985)}]{maLocalizedSuperconductors1985}%
  \BibitemOpen
  \bibfield  {author} {\bibinfo {author} {\bibfnamefont {M.}~\bibnamefont
  {Ma}}\ and\ \bibinfo {author} {\bibfnamefont {P.~A.}\ \bibnamefont {Lee}},\
  }\href {\doibase 10.1103/PhysRevB.32.5658} {\bibfield  {journal} {\bibinfo
  {journal} {Physical Review B}\ }\textbf {\bibinfo {volume} {32}},\ \bibinfo
  {pages} {5658} (\bibinfo {year} {1985})},\ \bibinfo {note} {publisher:
  American Physical Society}\BibitemShut {NoStop}%
\bibitem [{\citenamefont {Cea}\ \emph {et~al.}(2014)\citenamefont {Cea},
  \citenamefont {Bucheli}, \citenamefont {Seibold}, \citenamefont {Benfatto},
  \citenamefont {Lorenzana},\ and\ \citenamefont
  {Castellani}}]{ceaOpticalExcitationPhase2014}%
  \BibitemOpen
  \bibfield  {author} {\bibinfo {author} {\bibfnamefont {T.}~\bibnamefont
  {Cea}}, \bibinfo {author} {\bibfnamefont {D.}~\bibnamefont {Bucheli}},
  \bibinfo {author} {\bibfnamefont {G.}~\bibnamefont {Seibold}}, \bibinfo
  {author} {\bibfnamefont {L.}~\bibnamefont {Benfatto}}, \bibinfo {author}
  {\bibfnamefont {J.}~\bibnamefont {Lorenzana}}, \ and\ \bibinfo {author}
  {\bibfnamefont {C.}~\bibnamefont {Castellani}},\ }\href {\doibase
  10.1103/PhysRevB.89.174506} {\bibfield  {journal} {\bibinfo  {journal}
  {Physical Review B}\ }\textbf {\bibinfo {volume} {89}},\ \bibinfo {pages}
  {174506} (\bibinfo {year} {2014})},\ \bibinfo {note} {publisher: American
  Physical Society}\BibitemShut {NoStop}%
\bibitem [{\citenamefont {Ioffe}\ and\ \citenamefont
  {Mézard}(2010)}]{ioffeDisorderDrivenQuantumPhase2010}%
  \BibitemOpen
  \bibfield  {author} {\bibinfo {author} {\bibfnamefont {L.~B.}\ \bibnamefont
  {Ioffe}}\ and\ \bibinfo {author} {\bibfnamefont {M.}~\bibnamefont
  {Mézard}},\ }\href {\doibase 10.1103/PhysRevLett.105.037001} {\bibfield
  {journal} {\bibinfo  {journal} {Physical Review Letters}\ }\textbf {\bibinfo
  {volume} {105}},\ \bibinfo {pages} {037001} (\bibinfo {year} {2010})},\
  \bibinfo {note} {publisher: American Physical Society}\BibitemShut {NoStop}%
\bibitem [{\citenamefont {Lemarié}\ \emph {et~al.}(2013)\citenamefont
  {Lemarié}, \citenamefont {Kamlapure}, \citenamefont {Bucheli}, \citenamefont
  {Benfatto}, \citenamefont {Lorenzana}, \citenamefont {Seibold}, \citenamefont
  {Ganguli}, \citenamefont {Raychaudhuri},\ and\ \citenamefont
  {Castellani}}]{lemarieUniversalScalingOrderparameter2013}%
  \BibitemOpen
  \bibfield  {author} {\bibinfo {author} {\bibfnamefont {G.}~\bibnamefont
  {Lemarié}}, \bibinfo {author} {\bibfnamefont {A.}~\bibnamefont {Kamlapure}},
  \bibinfo {author} {\bibfnamefont {D.}~\bibnamefont {Bucheli}}, \bibinfo
  {author} {\bibfnamefont {L.}~\bibnamefont {Benfatto}}, \bibinfo {author}
  {\bibfnamefont {J.}~\bibnamefont {Lorenzana}}, \bibinfo {author}
  {\bibfnamefont {G.}~\bibnamefont {Seibold}}, \bibinfo {author} {\bibfnamefont
  {S.~C.}\ \bibnamefont {Ganguli}}, \bibinfo {author} {\bibfnamefont
  {P.}~\bibnamefont {Raychaudhuri}}, \ and\ \bibinfo {author} {\bibfnamefont
  {C.}~\bibnamefont {Castellani}},\ }\href {\doibase
  10.1103/PhysRevB.87.184509} {\bibfield  {journal} {\bibinfo  {journal}
  {Physical Review B}\ }\textbf {\bibinfo {volume} {87}},\ \bibinfo {pages}
  {184509} (\bibinfo {year} {2013})},\ \bibinfo {note} {publisher: American
  Physical Society}\BibitemShut {NoStop}%
\bibitem [{\citenamefont {{Ilaria Maccari}}\ \emph {et~al.}(2018)\citenamefont
  {{Ilaria Maccari}}, \citenamefont {{Lara Benfatto}},\ and\ \citenamefont
  {{Claudio Castellani}}}]{ilariamaccariBKTUniversalityClass2018}%
  \BibitemOpen
  \bibfield  {author} {\bibinfo {author} {\bibnamefont {{Ilaria Maccari}}},
  \bibinfo {author} {\bibnamefont {{Lara Benfatto}}}, \ and\ \bibinfo {author}
  {\bibnamefont {{Claudio Castellani}}},\ }\href {\doibase
  10.3390/condmat3010008} {\bibfield  {journal} {\bibinfo  {journal} {Condensed
  Matter}\ }\textbf {\bibinfo {volume} {3}},\ \bibinfo {pages} {8} (\bibinfo
  {year} {2018})}\BibitemShut {NoStop}%
\bibitem [{\citenamefont {Nelson}\ and\ \citenamefont
  {Kosterlitz}(1977)}]{nelsonUniversalJumpSuperfluid1977}%
  \BibitemOpen
  \bibfield  {author} {\bibinfo {author} {\bibfnamefont {D.~R.}\ \bibnamefont
  {Nelson}}\ and\ \bibinfo {author} {\bibfnamefont {J.~M.}\ \bibnamefont
  {Kosterlitz}},\ }\href {\doibase 10.1103/PhysRevLett.39.1201} {\bibfield
  {journal} {\bibinfo  {journal} {Physical Review Letters}\ }\textbf {\bibinfo
  {volume} {39}},\ \bibinfo {pages} {1201} (\bibinfo {year}
  {1977})}\BibitemShut {NoStop}%
\end{thebibliography}%

\end{document}